\documentclass[twocolumn,showpacs,prb]{revtex4-1}
\usepackage{graphicx}
\usepackage{dcolumn}
\usepackage{bm}
 
\def\bea{\begin{eqnarray}}
\def\eea{\end{eqnarray}}
\def\ben{\begin{equation}}
\def\een{\end{equation}}
\def\bit{\begin{itemize}}
\def\eit{\end{itemize}}
\input rotate
\begin{document}
\preprint{DIPC}
\title{Momentum-space finite-size corrections for Quantum-Monte-Carlo calculations}
 \author{R. Gaudoin$^{1}$, I.~G.~Gurtubay$^{2,3}$, and
J.~M.~Pitarke$^{2,4}$}
\affiliation{
$^{1}$Faculty of Physics, University of Vienna, and Center for Computational Materials Science, Sensengasse 8/12, A-1090 Vienna, Austria\\
$^{2}$Materia Kondentsatuaren Fisika Saila, Zientzia eta Teknologia Fakultatea, Euskal Herriko Unibertsitatea, UPV/EHU, 644 Posta kutxatila, E-48080 Bilbo, Basque Country, Spain\\
$^{3}$Donostia International Physics Center (DIPC), E-20018 Donostia,
Basque Country, Spain\\
$^{4}$CIC nanoGUNE Consolider and Centro F\'\i sica Materiales (CSIC-UPV/EHU), Tolosa Hiribidea 76,
E-20018 Donostia, Basque Country, Spain
}
\date{\today}

\begin{abstract}
Extended solids are frequently simulated as finite systems with
periodic boundary conditions, which due to the long-range nature of
the Coulomb interaction may lead to slowly decaying finite-size
errors. In the case of Quantum-Monte-Carlo simulations, which are
based on real space, both real-space and momentum-space
solutions to this problem exist. Here, we describe a hybrid method which using
real-space data models the spherically averaged structure factor in
momentum space. We show that (i) by integration our hybrid
method exactly maps onto the real-space model periodic 
Coulomb-interaction (MPC) method and (ii) therefore our method
combines the best of both worlds (real-space and momentum-space). One
can use known momentum-resolved behavior to improve convergence where
MPC fails (e.g., at surface-like systems). In contrast to pure
momentum-space methods, our method only deals with a simple
single-valued function and, hence, better lends itself to interpolation with
exact small-momentum data as no directional information is needed. By
virtue of integration, the resulting finite-size corrections can
be written as an addition to MPC.
\end{abstract}
\maketitle

\section{Introduction}
The issue of finite-size corrections in Quantum-Monte-Carlo (QMC)
calculations has recently been attracting considerable
attention.\cite{MPC1,CC,offdiag,DRUM,GP1} As QMC calculations of solids need to be
carried out within a supercell, the Coulomb interaction is typically
replaced by the so-called Ewald interaction that is compatible with
the supercell geometry.\cite{rmp} This, in turn, is equivalent to
dealing with an infinite system with a periodically repeated
exchange-correlation (xc) hole. In other words, using the Ewald
interaction includes the spurious effective interaction of an electron
with its periodically repeated xc hole. One solution to this drawback
- the real-space solution - is to use QMC data at length scales smaller than
the supercell size where QMC is expected to be accurate and
substitute the missing terms implicitly or explicitly. E.g., the Model
Periodic Coulomb-interaction (MPC) method\cite{MPC1} deals with the
periodically repeated xc hole by using the bare Coulomb term within
the supercell and assuming that beyond the supercell an electron only
feels the Hartree potential with no further correlations. This
represents a good approximation in a bulk solid, where the xc hole
decays rapidly. In contrast, Chiesa {\it al.}\cite{CC} traced the
Ewald error back to an integration error and then added back the missing contribution,
yielding a momentum-space solution.

In a recent paper,\cite{GP1} the spherically-averaged structure factor
$S_k$ was introduced in order to analyze and reduce Coulomb
finite-size effects. This ($S_k$) represents a natural quantity to
study finite size errors: On the one hand, it is a simple
one-dimensional function which via integration yields the
interaction energy and, on the other hand, it naturally orders the QMC
data according to length scale. Although QMC can, in principle, model
correlations well at short length scales, it is bound to
fail at larger scales due to the fact that finite simulation cells must
be used.

So far we have only touched on the issue of finite-size corrections
for the interaction energy. The information to evaluate this is
contained in the spherical average of the diagonal terms of the
two-particle density matrix (the two particle density) and hence $S_k$,
which contains the same information, suffices. However, other
quantities may be computed using QMC which also suffer from
finite-size errors. E.g. Chiesa {\it al.}\cite{CC} also deal with corrections
to the finite-size errors in the kinetic energy but since the kinetic energy 
needs information that goes beyond the diagonal terms of the density matrix
such an analysis is beyond the scope of this paper which is based solely 
on the information in $S_k$. Similarly, finite-size corrections to
the momentum distribution\cite{offdiag} cannot readily be done using
just $S_k$.

The aim of this paper is to show that modelling $S_k$ with the MPC
ansatz yields a momentum-resolved MPC. This allows for an intuitive
analysis of finite-size errors and for an improvement of MPC whenever
MPC fails, such as in surface-like systems. First of all, in Section II we discuss $S_k$ in
the context of QMC and apply the MPC ansatz, thereby showing that via
integration over $k$ our method exactly maps onto MPC. In Section III,
we first look at a non-interacting uniform electron gas in order to (i) see how MPC breaks
down in this case and (ii) learn how to overcome the limitation of MPC
by using our hybrid analysis. We also report Diffusion Monte Carlo (DMC)
calculations of the structure factor of an interacting uniform electron gas,
which yields an intuitive understanding of the MPC in interacting systems.
This then leads us to an easy way to understand the existing difficulties that arise in the case of a slab
geometry,\cite{2dqmc,surface} and we propose a simple method to
address this difficult problem. We round off the paper with a summary
and conclusions. We use atomic units throughout ($\hbar=e^2=m_e=1$).

\section{Modelling the exchange correlation hole in momentum space}
\subsection{The interaction energy}
The interaction energy $U^{int}$ of an arbitrary system of many interacting electrons can be expressed as follows
\begin{equation}
U^{int}=U_H+U_{xc},
\label{eq1}
\end{equation}
where $U_{H}$ represents the Hartree energy (which in the case of an infinite jellium model is exactly cancelled by the presence of the positive background) 
\begin{equation}
{U_{H}={e^2\over 2}\int d{\bf r}n({\bf r})\int d{\bf r}'\,{n({\bf r}')\over|{\bf r}-{\bf r}'|}},
\label{eq2}
\end{equation}
and $U_{xc}$ represents the so-called xc interaction energy corresponding to the attractive interaction between each electron and its own xc hole:
\begin{equation}
{U_{xc}={e^2\over 2}\int d{\bf r}n({\bf r})\int d{\bf r}'\,{n_{xc}({\bf r},{\bf r}')\over|{\bf r}-{\bf r}'|}}.
\label{eq3}
\end{equation}
Here, $n({\bf r})$ is the electron density and $n_{xc}({\bf r},{\bf r}')$ represents the xc-hole density of an electron at ${\bf r}$.
For brevity, we shall also define a {\it reduced} electron density $n_{red}({\bf r},{\bf r}')$, which represents the
electron density at ${\bf r}'$ seen by a given electron at ${\bf r}$ in the presence of (hence, reduced by) its xc hole: 
\begin{equation}
\label{nxcdef}
n_{red}({\bf r},{\bf r}')=n({\bf r}') + n_{xc}({\bf r},{\bf r}')=
\frac{n^2({\bf r},{\bf r}')}{n({\bf r})},
\label{reduced}
\end{equation}
where $n^2({\bf r},{\bf r}')$ is the so-called two-particle density.\cite{fetter} Note that $n_{xc}({\bf r},{\bf r}') <0$.

\subsection{Model periodic Coulomb interaction (MPC)}
The only periodic solution of Poisson's equation for a periodic array
of charges is the so-called Ewald interaction, which is of the Coulomb
form $1/r$ only in the limit of an infinitely large simulation
cell. However, while the Hartree energy is given correctly by this
Ewald interaction, the part of the electron-electron energy coming
from the interaction of electrons with their own xc hole yields a
spurious contribution that is due to the interaction of an electron
with its periodically repeated xc hole.  This drawback was solved in
Ref.~\onlinecite{MPC1} by replacing the Ewald interaction by a model
periodic Coulomb interaction that yields an interaction energy
consisting of the sum of two terms: The Hartree energy $U_H$ calculated with
the Ewald interaction, and the beyond-Hartree xc-energy $U_{xc}$
calculated with a cutoff Coulomb interaction using the minimum image convention,
 i.e., translating coordinates of electron pairs such that ${\bf r}-{\bf r'}$ lies within the simulation cell.
Hence, the MPC interaction energy $U^{int}$ is obtained by simply replacing the true {\it reduced} electron
density $n_{red}({\bf r},{\bf r}')$ of Eq.~(\ref{reduced}) by the MPC {\it reduced} electron density
$n^{MPC}_{red}({\bf r},{\bf r}')$ of the form displayed in Fig.~1 by the thick solid line marked ``MPC''.
\begin{figure}
\centering
\includegraphics[width=0.48\textwidth]{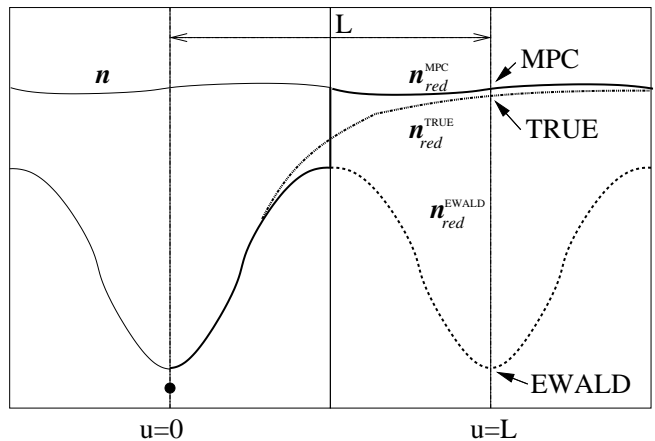}
\caption{The {\it reduced} electron density $n_{red}({\bf r},{\bf r}')$ 
seen by an electron at ${\bf r}$, as a function of $u=|{\bf r}-{\bf r}'|$.
In the absence of exchange and correlation, this
would equal the electron density $n({\bf r}')$ (thin solid line) which in
the case of a uniform electron gas would be constant. In the
presence of exchange and correlation, the true {\it reduced}
electron density $n_{red}({\bf r},{\bf r}')$ is of the form represented by the thick line marked ``TRUE'', which for a supercell geometry (of linear dimension $L$)
becomes the {\it reduced} Ewald density represented by the thick dotted line marked ``EWALD".
The MPC interaction energy $U^{int}$ is obtained by using a cutoff {\it reduced} electron density
(thick solid line marked ``MPC'') that coincides with (i) the {\it reduced} Ewald 
electron density within the simulation cell and (ii) the electron density $n({\bf r}')$
beyond the simulation cell.}
\label{f1}
\end{figure}

\subsection{Spherically averaged structure factor $S_k$}
Starting with the xc-hole density $n_{xc}({\bf r},{\bf r}')$ at ${\bf r}'$ around an
electron at ${\bf r}$, one finds the
following momentum-resolved form of the xc interaction energy of
Eq.~(\ref{eq3}):\cite{sphav}
\begin{equation}
U_{xc}={N\over\pi}\int (S_k-1)\, dk,
\label{eq4}
\end{equation}
where $S_k$ is the spherical average of the diagonal structure factor
$S_{{\bf k},{\bf k}'}$:~\cite{GP1}
\begin{equation}
\label{SF_base_eq}
S_k=1+{1\over N}\int d{\bf r}\,n({\bf r})\,\int d{\bf r'}\,
{\sin(ku)\over ku}\,n_{xc}({\bf r},{\bf r'}),
\end{equation} 
and $u=|{\bf r}-{\bf r}'|$.

Equation~(\ref{eq4}) represents a general expression for the xc interaction energy, which is formally exact not only for homogeneous media but also for an arbitrary inhomogeneous many-electron system.

\subsection{Monte-Carlo (MC) sampling of the structure factor}
When performing MC sampling on a function
$f({\bf r},{\bf r}')$ such as the Coulomb potential
$1/|{\bf r}-{\bf r}'|$, what we are actually calculating is
\begin{equation}
\left\langle \sum_{i\neq j} f({\bf r_i},{\bf r_j})\right\rangle_{MC}
=
\int d{\bf r}\,n({\bf r}) \int d{\bf r'}\,n^2({\bf r},{\bf r'})
f({\bf r},{\bf r}'),
\end{equation}
where the minimum image convention is used. In conjunction with Eq. (\ref{nxcdef}), this yields
the MC sampling of the spherically averaged structure factor of Eq. (\ref{SF_base_eq}):
\begin{equation}
S_k=1+{1\over N}
\left\langle\sum_{i\neq j}
{\sin(k|{\bf r_i}-{\bf r_j}|)\over k|{\bf r_i}-{\bf r_j}|}\right\rangle_{MC}
-S^{H}_k,
\end{equation} 
here $S^{H}_k$ being the Hartree contribution:
\begin{equation}
S^H_k={1\over N}\int_{SC} d{\bf r}\, d{\bf u}\, n({\bf r})\, n({\bf r}-{\bf u}) 
{\sin(ku)\over ku},
\end{equation}
where again the minimum image convention is being employed, i.e., the integrals
are carried out over the simulation cell (SC). The densities $n({\bf r})$ here are standard
MC electron densities.

\subsection{MPC from $S_k$}
If one applies the MPC ansatz in the calculation of the spherically
averaged structure factor $S_k$, by simply assuming (as in Fig. 1)
that beyond the supercell correlations are only due to variations in the density, then Eq.~(\ref{eq4}) yields exactly
the MPC interaction energy of Ref.~\onlinecite{MPC1}. This can be seen
by introducing Eq.~(\ref{SF_base_eq}) into Eq.~(\ref{eq4}) and performing
the $k$ integration. Hence, the momentum-space based method of
Ref.~\onlinecite{GP1} reduces, under the MPC ansatz, to what we may call a momentum-resolved MPC.

The MPC ansatz implies a finite extent of the xc hole, which in turn results in a quadratic
behavior of $S_k$ as $k\to 0$. In the case of finite systems (e.g.,
atoms, molecules, and clusters) and bulk solids, this quadratic
behavior of the structure factor is qualitatively correct because of
the short range of the xc hole in those systems. Hence, in those
systems, as long as the simulation cell is sufficiently large for the
spherically averaged structure factor at the cutoff momentum
$k_c\sim 1/L$ to already be in or close to the asymptotic low-$k$ behavior, the MPC
ansatz yields accurate results: Since in those systems the true $S_k$ is an essentially quadratic
function as $k\to 0$, constrained by $S_{k=0}=0$, and the same constraints hold
for $S^{MPC}_k$, only higher-order terms contribute to any residual error.

There are some caveats, however. The non-interacting uniform electron
gas as well as semi-infinite systems contain a linear contribution to $S_k$ at $k\to 0$ (due to the presence of a xc
hole that is not negligible even at large distances), which causes the
failure of the MPC scheme. Nevertheless, in the framework of our
momentum-resolved approach there is room for improvement over MPC, since one has
 flexibility to go beyond the 
MPC ansatz by replacing the low-$k$ structure factor $S^{MPC}_k$ by 
its known correct value.

\section{The MPC structure factor in practice}

\subsection{The non-interacting uniform electron gas}
\begin{figure}
\centering
\includegraphics[width=0.48\textwidth]{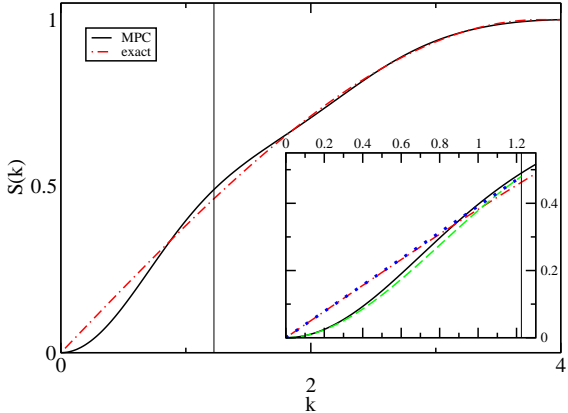}
\caption{(Color online)
The spherically averaged structure factor $S_k$ of a
  non-interacting (Hartree-Fock) uniform electron gas with an
  electron-density parameter $r_s=1$ and 54 electrons in a
  face-centered-cubic (fcc) simulation cell.
The exact HF structure factor $S_k$ is represented by a dashed-dotted red line and the MPC result by a solid black line. These two curves are also plotted in the inset, but now together with
Eq.~(\ref{smod2}) (nearly atop the exact HF structure factor), represented by a blue dotted line, and the modelling of
Eq.~(\ref{smod1}) (nearly atop the MPC HF structure factor), represented by a green dashed line. The difference
between the corresponding integrals represents  the correction to the MPC HF data. The vertical line indicates the cutoff $k_c$.}
\label{f2}
\end{figure}

The non-interacting (Hartree-Fock) uniform electron gas was dealt with
extensively in Ref.~\onlinecite{GP1}. Here we briefly discuss this
system, with the aim of now introducing a correction term. The exact
Hartree-Fock (HF) structure factor $S_k$ of a uniform electron gas, which
is easily derived analytically,\cite{HFSF} is shown in Fig.~\ref{f2}
(dashed-dotted red line) together with the result we obtain by using the MPC
ansatz (solid black line). This figure clearly shows that the MPC result is
in qualitative error at low $k$, due to the fact that the actual HF
structure factor does not exhibit a quadratic behavior in the limit as
$k\to 0$ but a linear behavior instead. Our momentum-resolved
technique, however, has room for improvement, by going beyond the MPC
ansatz.

Beyond a system-size dependent momentum cutoff $k_c$ (i.e., at
$k>k_c$), the QMC electron correlation and hence structure factor are
expected to be accurate. The momentum cutoff $k_c$ should be of
the order of the inverse of the characteristic length $L$ of the
supercell. A simple way to model a correction proceeds as follows:
After choosing an appropriate cutoff $k_c$ (vertical line of Fig.~\ref{f2}), we model the {\it corrected}
HF structure factor as a straight line between $S_{k=0}=0$ and $S^{MPC}_{k_c}$ and the
MPC structure factor as a quadratic curve between the same points.  The correction
to MPC is then given by the area between these two curves. The dotted blue line and the dashed green line of Fig.~\ref{f2}
illustrate this modelling. More realistic models are of course feasible. E.g., one could take into account
derivatives of $S_k$ at $k=0$ or $k=k_c$, etc.    

Hence, as long as we are interested in the xc
interaction energy $U_{xc}$, there is no need to actually evaluate the
spherically averaged structure factor for all $k$. In fact, if we assume that $k_c$ is already in the 
asymptotic regime one can base the entire correction on the value of $k_c$ and the 
asymptotic form of the structure factor.
Let us assume the MPC structure
factor below $k_c$ is of the form 
\begin{equation}
\label{smod1}
S^{MPC}_k = \beta k^2 + \gamma k^3  \ , 
\end{equation}
while the true functional dependence (in the asymptotic region) ought to be 
\begin{equation}
\label{smod2}
S_k = \alpha k \ , 
\end{equation}
$\alpha$ coming from the linear term in the HF $S_k$.\cite{HFSF}
At the cutoff, both should of course coincide:
\begin{equation}
\alpha
k_c=\beta k_c^2 +\gamma k_c^3\ , 
\end{equation} 
as should their derivatives:
\begin{equation}
\alpha
=2\beta k_c +3\gamma k_c^2,
\end{equation} 
which yields $\beta=2\alpha/k_c$ and  $\gamma=-\alpha/k^2_c$.

In this model, the {\it corrected} structure factor $S_k$ only
differs from the MPC structure factor at momenta between zero and $k_c$, so the correction in $U_{xc}$ can
easily be evaluated using Eq.~(\ref{eq4}); it turns out to be: 
\begin{equation}
\label{hfcorr}
\Delta U_{xc}=U_{xc}^{\rm corrected}-U_{xc}^{MPC}
= \frac{1}{12}\alpha k_c^2.
\end{equation}
With $k_c\sim{1}/{L}$, we immediately get that the MPC error scales as
$\sim{1}/{L^2}\sim{1}/{V^{2/3}}$.

We now also see how limiting the constraint of a finite cutoff when
using an MPC like analysis actually is: The quadratic model HF structure
factor of Eq.~(\ref{smod1}) (see the dashed green line of Fig.~\ref{f2}) and
the actual MPC HF structure factor (solid black lines of Fig.~\ref{f2}) are nearly identical and
quite different from the exact HF structure factor (dashed-dotted red line and dotted blue line of Fig.~\ref{f2}).
This is despite the model MPC HF structure factor only using the asymptotic
quadratic shape and the value of the MPC HF structure factor at the cutoff where it ought to coincide with the exact HF structure factor.

As pointed out above, so far this represents a crude way of devising the finite-size
correction to the MPC result. More complex functional forms can be
chosen, by using the structure factor or even its derivatives at $k=0$
or $k\neq 0$ to estimate the corresponding parameters. Analytic
integration would then yield a more accurate correction term.

\subsection{The interacting uniform electron gas}
In Ref.~\onlinecite{GP1}, Variational Monte Carlo (VMC) calculations
of the structure factor $S_k$ and the xc interaction energy $U_{xc}$
of an interacting uniform electron gas were reported. Here we report DMC
calculations of these quantities, as
obtained by following the Hellmann-Feynman sampling introduced in
Ref.~\onlinecite{GP2}.

\begin{figure}
\centering
\includegraphics[width=0.48\textwidth]{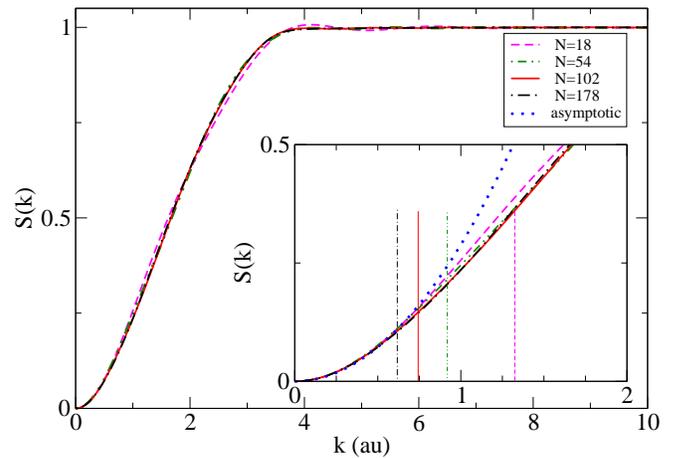}
\caption{(Color online)
DMC spherically averaged structure factor $S_k$ of an interacting uniform electron gas of $r_s=1$
(in an fcc simulation cell with $N$=18, 54, 102, and 178, $N$ being the number of electrons), at the MPC level. The thick dotted blue line marked ``asymptotic" represents the true structure factor of a uniform electron gas at $k\to 0$: $S_k=k^2/2\omega_p$.\cite{notelast} The vertical lines represent the cutoff $k_c=1/L$ for $N$=18, 54, 102, and 178.}
\label{f3}
\end{figure}

Figure~\ref{f3} exhibits the DMC structure factor $S_k$ of an interacting uniform electron gas of $r_s=1$, as obtained by following the
MPC {\it ansatz} for various values of $N$ (the number of electrons in the simulation cell). As pointed out in Ref.~\onlinecite{GP1},
the {\it true} interacting structure factor $S_k$ is quadratic at $k\to 0$.\cite{notelast} Fig.~\ref{f3} shows that our MPC DMC structure
factor nicely reproduces this low-$k$ limit (thick dotted blue line marked ``asymptotic"). The MPC DMC structure factor of Fig.~\ref{f3} could be improved by using at low wavevectors the structure factor that one can obtain numerically in the random-phase approximation (RPA), which is known to be accurate in the low-$k$ regime.

\subsection {Semi-infinite electron gas}
It is well known that in the case of a semi-infinite electron gas the
correct behavior of the structure factor $S_k$ at low $k$ consists of
a linear term accounting for the surface contribution that is
augmented by the usual quadratic bulk term. For a finite
simulation cell of width $L_z$ and surface area $A$, one
finds (in the long-wavelength limit $k\to 0$):\cite{slab}
\begin{equation}
S_k=\alpha k+\beta k^2,
\label{linear}
\end{equation} 
where now
\begin{equation}
\alpha=\frac{\pi}{L_z}\left[1/\omega_p-1/2\omega_s\right]
\label{alpha}
\end{equation}
and
\begin{equation}
\beta=1/2\omega_p.
\end{equation}
Here, $\beta$ represents the usual bulk term,\cite{notelast} and $\alpha$ follows from Eqs.~(2.13) and (3.34) of Ref.~\onlinecite{slab},
$\omega_s=\omega_p/\sqrt{2}$ being the surface-plasmon energy. As the system gets larger ($L_z\to\infty$) the relative surface contribution shrinks, as expected. Nonetheless, surface energies, which have no contribution from the bulk part of the structure factor, are entirely dominated by the
{\it linear} surface-contribution $\alpha k$. 

The MPC ansatz will never yield the linear term of
Eq.~(\ref{linear}). In order to solve this shortcoming, one can
perform a simple analysis similar to the one leading to Eq.~(\ref{hfcorr}), now
employing the linear surface term of Eq.~(\ref{alpha}). This would yield an MPC error that scales as
$\sim{1}/{V}$. However, for this to make sense one needs a
well defined cutoff $k_c$ below which the quadratic MPC behavior
should be replaced by the correct linear behavior of
Eq.~(\ref{linear}) and above which the MPC structure factor is
essentially exact. Nevertheless, in the case of realistic
calculations~\cite{surface} of surface energies where the surface area
needs to be varied for a fixed slab width and vice versa, finding a
well-defined cutoff $k_c$ might not be possible, so that one would
need to explore more complex functional forms of the structure factor
involving the MPC structure factor itself and possibly its derivatives at 
one or more values of $k$, such as $k=k_c$. In conjunction with the known surface
contribution $\alpha$ of Eq.~(\ref{alpha}), such schemes could correct
significantly the existing MPC surface calculations.

\subsection{Implicit correction to the Coulomb kernel}

An interesting observation is that any correction to the structure factor implies a
correction $\Delta V^c$ to the Coulomb potential $V^C=1/u$: For
$0<k<k_c\, \, $, $S_k^{MPC}$ is given by evaluating
the kernel $\sin(ku)/ku$ entering Eq.~(\ref{SF_base_eq}). 
Instead, sampling the quantity
\begin{equation}
\label{dker}
\left( {S_k^{cor.}-1\over{
S_k^{MPC}-1}}\right){\sin(ku)\over ku}
\end{equation}
gives $S_k^{cor.}$, the finite-size corrected structure
factor. This will differ from $S_k^{MPC}$ only between $k=0$ and $k=k_c$; the fraction
in Eq. (\ref{dker}) is therefore well-defined.
But the integral of the kernel, $(2/\pi)\int_0^{k_c}\sin(ku)/ku=1/u$, is precisely the implied Coulomb potential $V^C$.
Changing the kernel thus changes the implicit
Coulomb potential:
\begin{equation}
\label{vcor}
\Delta V^c={2\over\pi}\int^{k_c}_0 dk \left( {S_k^{cor.}-S_k^{MPC}\over{
    S_k^{MPC}-1}}\right){\sin(ku)\over ku}.
\end{equation}
This is a short-range correction in the sense 
that when performing the MC run $u$ is effectively restricted to within
the simulation cell. On the other hand, the $k$ in the integral is smaller 
than $k_c$, which is inversely proportional to the simulation-cell size, and so
$ku$ entering the $\sin$ function in Eq. (\ref{vcor}) is generally below $1$.

In the case of the model of Eqs.~(\ref{smod1}) and (\ref{smod2}) we get, for example,
\begin{equation}
\Delta V^c={2\over \pi u}\int^{k_c}_0 dk \left( 
{\alpha[1-{2k\over k_f} + {k^2\over k^2_f}]
\over{-1+\alpha{2k^2\over k_f}-\alpha {k^3\over k^2_f}
}}\right) 
\sin(ku).
\end{equation}
$S_k^H$  must then also be 
evaluated using $\Delta V^C$, as it is used in $S_k$ via $n_{xc}$.
In contrast, $U_H$ continues to be the standard Ewald
Hartree energy.
 
\section{Summary and conclusions}
First of all, we have demonstrated that modelling the spherically
averaged structure factor $S_k$ with the MPC ansatz (momentum-resolved MPC)
yields exactly, after integration, the MPC interaction energy of Ref.~\onlinecite{MPC1}. 
This allows us to see explicitly that in the case of solids and finite systems
MPC improves convergence considerably (over the more traditional Ewald scheme), due
to the fact that for such systems MPC yields the correct quadratic behavior of $S_k$ as $k\to 0$.
We also see explicitly how the MPC ansatz breaks down in the case of the non-interacting
uniform electron gas and, in general, in the case of all systems exhibiting a similar
pathology, i.e., an {\it extended} xc hole (at surfaces, for example),
where the leading term of $S_k$ at $k=0$ is proportional to $k$.

As the explicit $k$ dependence of $S_k$ at low wave vectors
can usually be known, one can look at QMC systems at different length scales separately enabling us
to analyze the xc interaction energy at different length scales and to
derive a correction term. On integration, this term yields a correction to
QMC calculations that are based on the model periodic-Coulomb interaction MPC.

\acknowledgments This work has been supported by the Basque
Unibertsitate eta Ikerketa Saila (Grant No. GIC07IT36607) and the Spanish Ministerio de Ciencia e Innovaci\'on (Grants No. FIS2009-09631 and No. CSD2006-53).


\begin{thebibliography}{100}
\bibitem{sphav} Note that only the spherically averaged xc hole enters
  here, which is due to the fact that the Coulomb interaction is
  spherically symmetrical and, hence, the angular information of the
  xc hole drops out even for inhomogeneous systems.
\bibitem{notelast} The true structure factor of a uniform electron gas
  at $k\to 0$ is $S_k=k^2/2\omega_p$; here, $\omega_p=(4\pi n)^{1/2}$
  represents the bulk-plasmon energy, $n$ being the electron density.
\end{thebibliography}
\end{document}